\documentclass[josab,twocolumn]{revtex4-1}
\usepackage{graphics}
\usepackage{color}
\begin{document}
\title{Monte Carlo Studies of the Intrinsic Second Hyperpolarizability}
\author{Shoresh Shafei}
\author {Mark C. Kuzyk*}
\author{Mark G. Kuzyk}
\address{Department of Physics and Astronomy, Washington State University, Pullman, Washington 99164-2814}
\address{*Current address: Department of Physics, University of Oregon, Eugene, Oregon 97403}
\address{Corresponding author: kuz@wsu.edu}

%\date{\today}
\begin{abstract}
 The first and second order hyperpolarizabilities have been extensively studied to identify universal properties near the fundamental limit.  Here, we employ the Monte Carlo method to study the fundamental limit of the second hyperpolarizability.  As it was found for the first hyperpolarizability, the largest values of the second hyperpolarizability approaches the calculated fundamental limit.  The character of transition moments and energies of the energy eigenstates are investigated near the second hyperpolarizability's upper bounds using the missing state analysis, which assesses the role of each pair of states in their contribution.  In agreement with the three-level ansatz, our results indicate that only three states (ground and two excited states) dominate when the second hyperpolarizability is near the limit.
\end{abstract}

\maketitle

\section{Introduction}
In 2000, the fundamental limit of the off-resonant electronic first hyperpolarizability and second hyperpolarizability were calculated in an attempt to understand constraints imposed by quantum mechanics \cite{kuzyk00.01,kuzyk00.02,kuzyk01.01}. (Note that for the rest of this paper, we say hyperpolarizability when referring to the first hyperpolarizability.) Using Thomas-Kuhn sum rules and assuming that when the hyperpolarizability $\beta$ is optimized only three states (including the ground state) contribute, an upper limit of $\beta$ was found to be given by,
\begin{equation}\label{betamax}
\beta_{max} = 3^{1/4} \left(\frac{e \hbar}{\sqrt{m}}\right)^3 \left(\frac{N^{3/2}}{E_{10}^{7/2}}\right) ,
\end{equation}
where $-e$ is charge of the electron, $\hbar$ is Planck's constant, $N$ denotes the number of electrons in the atom or molecule and $E_{10}$ is the energy difference between the first excited state and ground state.  In conjugated systems, it is known that only the $\pi$-electrons contribute to the nonlinear-optical response, so the effective number of electrons, $N=N_{eff}$ is used instead of the total number.\cite{zhou08.01,kuzyk09.01}

A comparison the largest experimentally measured second order molecular susceptibilities with the fundamental limit reveals a large gap between the two. Measurements have never crossed the limit and are typically well below it.  Until about 2005, all molecules ever measured mysteriously fell a factor of about 30 below the limit,\cite{kuzyk03.01,kuzyk03.02} until a new class of molecules were reported by Kang and coworkers.\cite{Kang05.01,Kang07.01}

In early studies,\cite{tripa04.01} several experimental tools such as linear spectroscopy, Raman spectroscopy, hyper-Rayleigh scattering and Stark spectroscopy were used to investigate the possible reasons for the gap between theory and experiment. There it was explained that the gap is mainly due to the unfavorable arrangement of excited state energies.

In an effort to understand how one might make a real quantum system that has a hyperpolarizability near the limit, Zhou and coworkers applied a numerical approach to determine the shape of the potential energy function that yields the largest value of the hyperpolarizability\cite{zhou06.01, zhou07.02}.  The potentials were chosen among a wide range of possibilities including polynomials, power laws and piecewise continuous functions.  The best hyperpolarizabilities found were given by $\beta \approx 0.7\beta_{max}$.  Thus, a new but smaller gap appeared between numerical simulations and the fundamental limit.  These studies suggested that the concept of conjugation of modulation could be used as a new design paradigm for making better molecules that was later experimentally verified with one-dimensional molecules that broke above the apparent limit.\cite{perez07.01,perez09.01}

Those numerical optimization studies showed the first hint of the appearance of universal properties of quantum systems that approach the fundamental limit.  In all cases, independent of the starting potential, a large range of optimized systems were found to have (1) $\beta \approx 0.7\beta_{max}$; (2) they are all well approximated by a three-level model; (3) the energy ratio between the first two excited states is always near 1/2; and, (4) the the normalized transition moment to the first excited state is about $0.75$.\cite{zhou06.01, zhou07.02} A broader set of studies found similar universal properties, for example, when considering the effects of geometry (i.e. the optimal arrangement of point nuclei)\cite{kuzyk06.02} and the effects of an externally applied electromagnetic field.\cite{watkins09.01}

Since the 30\% gap may be an indication of a flaw in the calculation of the fundamental limits, Kuzyk and Kuzyk (KK) focused on investigating the validity of the three level ansatz, the only assumption underlying limit theory.\cite{kuzyk08.01} The three level ansatz states that when the hyperpolarizability of a quantum system is at the fundamental limit, only three states contribute;\cite{kuzyk09.01} that is, the contributions from all other states is negligible.  Note that the ansatz does not claim the converse.  When three states are involved, the hyperpolarizability is not necessarily at the limit.  KK used Monte Carlo simulations to numerically generate a set of $\beta$ values to investigate if any of the data attains the predicted fundamental limit.  They found a continuous distribution of values up to the fundamental limit.  These studies verified the validity of the theory of limits and therefore indirectly verified the three level ansatz.

More interestingly, these results imply that a set of wavefunctions exist that are consistent with the sum rules, yet may not be derivable from typical Hamiltonians; that is, the sum rules may permit state vectors that are not derivable from the Schr\"{o}dinger equation.  This statement may seem contradictory since sum rules are directly derivable from the Schr\"{o}dinger equation; but, the sum rules apply under broader conditions, which we can understand as follows.

The sum rules are derived using the fact that $\left[ x, \left[x, H \right] \right] = \hbar^2/2m$ for a {\em typical Hamiltonian}.  However, it may be possible that $\left[ x, \left[x, H + A \right] \right] = \hbar^2/2m$ for an operator $A$, where $A$ can take a form that is never observed in molecular systems. As an example, it is straightforward to show that the unusual operator $A = f(x,y,z) p_x p_y p_z$ obeys the commutation relationship yet is of a form that is to our best knowledge never found in nature.  Other such operators can be invented.  Thus, while all standard Hamiltonians obey the sum rules, more exotic Hamiltonians can also obey the sum rules.

Sum rules have been used to determine unknown transition moments when fitting the dispersion data of the second harmonic hyperpolarizability,\cite{Hu2010.01} and for understanding length-scaling of the second hyperpolarizability.\cite{slepk04.01} Fundamental limits have also been used to investigate the essential ingredients required of small molecules to approach the fundamental limit.\cite{May05.01,May07.01} However, no numerical optimization studies have targeted the second hyperpolarizability.  In the present work, we focus on Monte Carlo simulations of $\gamma$, the second hyperpolarizability, to test the assumptions used in calculating the fundamental limits, to search for the broadest universal properties, and determine if this approach is capable of generating $\gamma$ values close to the limit.  Finally, we will analyze the number of states that contribute to $\gamma$ when it approaches the limit -- a first step in establishing the validity of the three-level ansatz for the second hyperpolarizability.

\section{Approach}
The second hyperpolarizability can be expressed using the dipole-free (DF) sum-over-states equation, which is equivalent to the traditional sum-over-states (SOS) expression when enough states are included in the sum.  The $xxxx$ tensor component of the off resonant dipole-free SOS expression of the second hyperpolarizability is given by,\cite{perez01.08}
\begin{widetext}
\begin{eqnarray} \label{gamma}
\gamma_{DF} &=& \frac{1}{8} \left( 2 {\sum_{n}^{\infty}}' {\sum_{m\neq n}^{\infty}}' {\sum_{l \neq n}^{\infty}}' \left\{ \frac{(2E_{m0}-E_{n0})(2E_{l0}-E_{n0})}{E_{n0}^{5}} - \frac{(2E_{l0}-E_{n0})}{E_{m0}E_{n0}^{3}} \right\} x_{0m}x_{mn}x_{nl}x_{l0} \nonumber \right. \\ &+& 2
{\sum_{n}^{\infty}}' {\sum_{m \neq n}^{\infty}}' {\sum_{l \neq m}^{\infty}}' \left\{\frac{1}{E_{l0}E_{m0}E_{n0}}-\frac{(2E_{l0}-E_{m0})}{E_{m0}^{3}E_{n0}}
\right\} x_{0l}x_{lm}x_{mn}x_{n0} - \left. {\sum_{m}^{\infty}}'{\sum_{n}^{\infty}}' \left\{\frac{1}{E_{m0}^{2}E_{n0}} +\frac{1}{E_{n0}^{2}E_{m0}} \right\}x_{0m}^{2}x_{0n}^{2} \right),\nonumber \\
\end{eqnarray}
\end{widetext}
where a prime indicates that the ground state is excluded in the summation.  $E_{ij} = E_i - E_j$ where $E_i$ and $E_j$ are energies of $i^{th}$ and $j^{th}$ states, respectively. $x_{ij}$ is the matrix element of position between states $i$ and $j$.  Note that Equation \ref{gamma} holds for a three-dimensional molecule where we have chosen the coordinate system such that $xxxx$ is the largest diagonal component of the second hyperpolarizability.

The Thomas-Kuhn sum rules are the direct consequence of the commutator of the Hamiltonian with the position operator, $\left[\left[H, x\right], x\right]$ where $H$ is a Hamiltonian of the form
\begin{equation}\label{H}
H(\mathbf{p},\mathbf{r}) = \frac{p^2}{2m} + V(\mathbf{r}).
\end{equation}
Here, $p$ is the momentum at position $\mathbf{r}$ and $V(\mathbf{r})$ is the potential energy of the particle. The Thomas-Kuhn sum rules for the non-relativistic Hamiltonian Equation \ref{H} are given by,
\begin{equation}\label{SR}
\sum_{n=0}^{\infty} \left[E_n - \frac{1}{2}(E_m + E_p)\right] x_{mn}x_{np} = \frac{\hbar^2 N}{2 m}\delta_{mp}
\end{equation}
where $\delta_{ij}$ is the Kronecker delta.  Equation \ref{SR} represents an infinite number of equations indexed by the integers $(m,p)$.  In the text that follows, we refer to a particular sum rule by stating the values $(m,p)$.  Note that Equation \ref{SR} holds for a broader set of multi-electron Hamiltonians including spin, externally applied electromagnetic fields, and electron correlations.\cite{kuzyk10.01}   Most of the cases can be described by the non-relativistic Hamiltonian. For the relativistic case, some modifications of Equation \ref{SR} are required.\cite{Goldman82.01, Leung86.01, cohen04.01}

The normalized form of the sum rules take the form
\begin{equation}\label{sumrules}
\sum_{n=0}^{\infty} \left[e_n - \frac{1}{2}(e_m + e_p)\right] \xi_{mn}\xi_{np} = \delta_{mp},
\end{equation}
where $e_i = E_{i0}/E_{10}$ can take on values from one to infinity for $i > 1$ (the ground state is labeled by $0$). Finally the normalized transition moments are defined as
\begin{equation}\label{transitionMoments}
\xi_{ij} = \frac{x_{ij}}{|x_{01}^{max}|}
\end{equation}
where
\begin{equation}\label{x01max}
|x_{01}^{max}|^2 =\frac{\hbar^2 N}{ 2mE_{10}}
\end{equation}
Equations \ref{x01max} and \ref{SR} imply that the largest possible transition moment from the ground state is $x_{01}$ because $E_{10} < E_{i0}$ for $i>1$.

The three level ansatz is central to calculating the fundamental limit of the second hyperpolarizability, $\gamma$.\cite{kuzyk00.02}  It states that when $\gamma$ is at the limit, only three states contribute. Using this ansatz and the sum rules given by Equation \ref{SR}, the second hyperpolarizability given by Equation \ref{gamma} is found to be constrained to the range,\cite{kuzyk00.02}
\begin{eqnarray}\label{gammamax}
-\frac{e^4 \hbar^4}{m^2}\left(\frac{N^2}{E_{10}^5}\right) \leq \gamma \leq 4\frac{e^4 \hbar^4}{m^2}\left(\frac{N^2}{E_{10}^5}\right).
\end{eqnarray}
Hence, defining the fundamental limit of the second hyperpolarizability as
\begin{eqnarray}\label{gammamax2}
\gamma_{max} = 4\frac{e^4 \hbar^4}{m^2}\left(\frac{N^2}{E_{10}^5}\right),
\end{eqnarray}
then
\begin{eqnarray}\label{gammamax3}
-\frac{1}{4} \leq \gamma_{int} \leq 1,
\end{eqnarray}
where $\gamma_{int} = \gamma/ \gamma_{max}$. Instead of studying $\gamma$ directly, we use the dimensionless intrinsic second hyperpolarizability because it takes into account simple scalings.\cite{kuzyk10.01} Hence, when $\gamma$ approaches $e^4 N^2\hbar^4/m^2 E_{10}^5$, $\gamma_{int}$ approaches unity.

The numerical procedure is the same as in previous studies of the hyperpolarizability, $\beta$:\cite{kuzyk08.01} For a specific number of states $s$, the energy values are picked arbitrarily and sorted in ascending order so that $ E_{0} < E_{1} < \ldots < E_{s-1}$ where $E_{0}$ is the energy of the ground state and $E_{i}$ is the energy of the $i^{th}$ excited state.  For simplicity and with no loss in generality, we shift the energies so that the ground state energy is zero. Since $e_{i} = E_{i0}/E_{10}$, then $e_{0}=0$ and $e_{1}=1$.

The next step is assigning transition moments that together with the energies are forced to be consistent with sum rules.  For a time-invariant Hamiltonian, it can be shown that the wavefunctions are real provided that there are no degeneracies.\cite{sakur94.01}  Our method is not well suited to situations where degenerate states are present. As such, we treat the case where transition moments are real and intentionally avoid degeneracies.  We are thus ignoring the class of problems in which (1) the Hamiltonian is not time invariant; and, (2) there are degeneracies.  These cases will be considered in future studies.

Since there are no dipole moments (the diagonal components of the transition moments) in $\gamma_{DF}$ given by Equation \ref{gamma}, we need only use the sum rules with $m=p$ in Equation \ref{sumrules} to get all of the required transition moments.  Starting with $(m,p)=(0,0)$, we get
\begin{equation}\label{m=p=0}
e_{10}|\xi_{10}|^2 + e_{20}|\xi_{20}|^2 + e_{30}|\xi_{30}|^2 + \ldots + e_{n0}|\xi_{n0}|^2 = 1.
\end{equation}
Since $e_{10} = e_1 - e_0 = 1$, $\xi_{01}$ can be randomly picked from the interval $ -1 < \xi_{01} <  1$. Then $\xi_{02}$ is obtained based on the fact that
\begin{equation}\label{sumrule00-1}
\sum_{n=2} e_{n0}\xi_{n0}^2 = 1 - e_{10}\xi_{10}^2,
\end{equation}
which implies
\begin{equation}\label{ineq1}
\xi_{20}^2 \leq \left( 1 - e_1 \xi_{10}^2 \right)/e_{20}.
\end{equation}
A random number $-1 \leq r \leq 1$ is used to get $\xi_{20}$ from Equation \ref{ineq1},
\begin{equation}\label{rule2}
\xi_{20} = r\sqrt{\left( 1 - e_1 \xi_{10}^2\right)/e_{20}}.
\end{equation}
For an s-state model the same method is used for all other transition moments, $\xi_{i0}$, except for $\xi_{s-1,0}$, which is directly determined from the sum rules.  For example, for a 10 state model,
\begin{equation}\label{xi10,0}
\xi_{10,0} = \sqrt{\left(1- e_1 \xi_{10}^2 - e_2 \xi_{20}^2-\ldots-e_9\xi_{90}^2\right)/e_{10,0}}.
\end{equation}

The remaining transition moments are calculated using the higher-order sum rules in sequence with $p = 1,2,3,\ldots$ in Equation \ref{sumrules}. Since $\xi_{ij}$ is assumed to be real for all states $i$ and $j$,
\begin{equation}\label{reverse}
\xi_{ij} = \xi_{ji}.
\end{equation}
The energy and transition moment values are inserted in normalized form into Equation \ref{gamma} to calculate $\gamma_{int}$. The normalized form is obtained by dividing Equation \ref{gamma} by Equation \ref{gammamax2} such that in Equation \ref{gamma}, $\gamma \rightarrow \gamma_{int}$, $E_{ij} \rightarrow e_{ij}$ and $x_{ij}\rightarrow \xi_{ij}$.  Note that one set of parameters obtained with the procedure given by Equations \ref{sumrule00-1} to Equation \ref{reverse} is called a single trial or single iteration.  To identify common properties of the second hyperpolarizability, hundreds of thousands of iterations are performed, and the distributions of the results are analyzed.

Before proceeding, we need to address the issue of the signs of the transition moments.  Recall that the diagonal sum rules suffice to fix all the transition moments. However, all the transition moments in the diagonal sum rules appear as a squared modulus, $|\xi_{ij}|^2$. Thus, the diagonal sum rules give no information about the sign of $x_{ij}$.  So one can choose the sign of each transition moment independently from the values of all others. The question arises whether or not the signs are constrained by the sum rules that were neglected.

To answer this question, we first refer to the non-diagonal sum rules for which $m\neq p$ in Equation \ref{sumrules}, which can be used to determine a condition on the signs by solving the equations for all different pairs of $m$ and $p$.  Considering the number of states, $s$, this approach demands solving $s$ coupled equations. Since this is time consuming, we confronted the problem using the simpler approach of randomly picking all transition moments to be positive or negative.  Neither $\beta$ nor $\gamma$, calculated from the resulting matrix elements and energies, exceeded the fundamental limit.  Thus we concluded that the sign of the transition moments can be picked randomly. We have studied the issue of truncated sum rules,\cite{kuzyk06.01} and in future, we will address the self-consistency between the sum rules when truncated.

The energies and transition moments are related via sum rules that impose a limit on all the transition moments. To express the limit mathematically we start with Equation \ref{sumrules} with $m=p$,
\begin{equation}\label{p=0-0}
\sum_{n=0}^{\infty}e_{np}|\xi_{np}|^2 = 1.
\end{equation}
For $p=0$ Equation \ref{p=0-0} yields
\begin{equation}\label{p=0-1}
e_{10}|\xi_{10}|^2 + e_{20}|\xi_{20}|^2 + e_{30}|\xi_{30}|^2 + \ldots + e_{n0}|\xi_{n0}|^2 = 1.
\end{equation}

Equation \ref{p=0-1} implies that the largest value of $e_{i0}|\xi_{i0}|^2$ is attained when all other terms of the type $e_{j0}|\xi_{j0}|^2$ (for $j\neq i$) are zero, i.e.
\begin{equation}\label{p=0-2}
e_{i0}|\xi_{i0}|^2 \leq 1 \hspace{1cm}  0 < i \leq n
\end{equation}
or
\begin{equation}\label{p=0-3}
-\left(\frac{1}{e_{i0}}\right)^{1/2} \leq \xi_{i0} \leq \left(\frac{1}{e_{i0}}\right)^{1/2}.
\end{equation}
$\xi_{i0}$ is largest when $e_{i0}$ is minimized and vice versa. Since based on our definition for energy spacing, when $i \geq 1$, we have $e_{i0} \geq 1$ then $\xi_{i0}$ is limited to
\begin{equation}\label{p=0-4}
-1 \leq \xi_{i0} \leq 1 \hspace{1em} \mbox{for} \hspace{1em} 0 < i \leq n .
\end{equation}

Following the same procedure for $p=1$ in Equation \ref{sumrules} results in
\begin{equation}\label{p=1-1}
e_{01}|\xi_{01}|^2 + e_{21}|\xi_{21}|^2 + e_{31}|\xi_{31}|^2 + \ldots = 1,
\end{equation}
whence
\begin{equation}\label{p=1-2}
 e_{21}|\xi_{21}|^2 + e_{31}|\xi_{31}|^2 + \ldots + e_{n1}|\xi_{n1}|^2 = 1 + e_{10}|\xi_{01}|^2.
\end{equation}
Here we have used the fact that $e_{01} =-e_{10}$. Substituting Equation \ref{p=0-2} into Equation \ref{p=1-2} gives an upper limit
\begin{equation}\label{p=1-3}
e_{21}|\xi_{21}|^2 + e_{31}|\xi_{31}|^2 + \ldots + e_{n1}|\xi_{n1}|^2 \leq 2.
\end{equation}
This implies that $e_{i1}|\xi_{i1}|^2 \leq 2$ (when $1 < i \leq n$) and is obtained when for all $j\neq i$, $e_{j1}|\xi_{j1}|^2 = 0$. Then
\begin{equation}\label{p=1-4}
e_{i1}|\xi_{i1}|^2 \leq 2 \hspace{1cm}  1 < i \leq n.
\end{equation}
or
\begin{equation}\label{p=1-5}
-\left(\frac{2}{e_{i1}}\right)^{1/2} \leq \xi_{i1} \leq \left(\frac{2}{e_{i1}}\right)^{1/2} \hspace{1em} \mbox{for} \hspace{1em} 1 < i \leq n .
\end{equation}

Using the same method for $p = 3,4,5,\ldots, n-1$, the constraint on the transition moments is found to be,
\begin{equation}\label{translimit}
-\frac{2^{\,p/2}}{e_{ip}^{1/2}} \leq \xi_{i,p} \leq \frac{2^{\,p/2}}{e_{ip}^{1/2}}  \hspace{1em} \mbox{for} \hspace{1em} p < i \leq n.
\end{equation}
Recall that we have assumed that the transition moments are real so $\xi_{ij} = \xi_{ji}$. The sum rules specify the range of validity of all non-diagonal transition moments according to Equation \ref{translimit}. The Monte Carlo method typically yields energy spacings that lead to transition moments that lie below unity. Equation \ref{translimit} is a general result that applied to quantum systems that are described by the Hamiltonian given by Equation \ref{H}.

\section{Results and Discussion}

In this section, we use Monte Carlo simulations to study the statistics of the distribution of the intrinsic second hyperpolarizability with the aim of understanding the physical properties of a quantum system that leads to a nonlinear response that approaches the fundamental limit. Correlations of the values of $\gamma_{int}$ with the {\em average} energy spacing and transition moments when $\gamma_{int}$ is at the limit will be used to investigate whether such systems behave in a way that suggests universal behavior.  Finally, we discuss the effect that pairs of states have on the second hyperpolarizability using a missing state analysis.

When energies are chosen randomly, on average they are equally spaced. Thus the energy spacing is similar to the eigenenergies of a harmonic oscillator.  To increase the domain of energy spacing that is probed by the Monte Carlo approach, we define an energy weighing factor $f$, such that
\begin{equation}\label{Espacing}
E_{i}\rightarrow E_{i}^f.
\end{equation}
For $f < 1$ the energies are denser at higher energies and for $f > 1$ the states are denser at lower energies.

\begin{figure}
\includegraphics{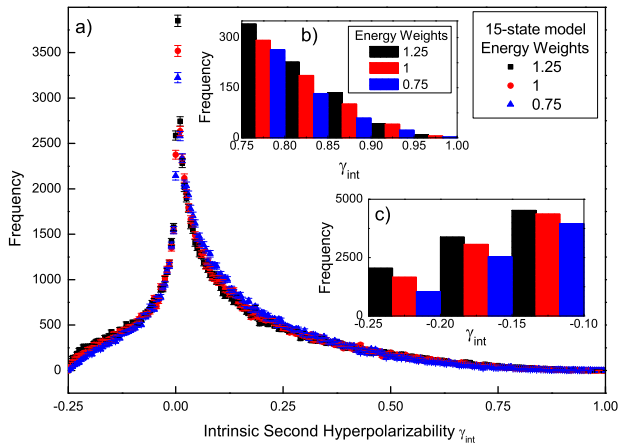}
\caption{(a)Distribution of simulated intrinsic second hyperpolarizabilities $\gamma_{int}$ for a 15-state model and weight factors of 0.75, 1, 1.25. The inset shows histograms of $\gamma_{int}$ near the limits. The bin size for the main figure is 0.003; and, for Inset b and Inset c are 0.05. The error bars are defined as the square root of the frequency.}\label{fig:15}
\end{figure}
Figure \ref{fig:15}a shows the distribution of the second hyperpolarizability for a 15-state model for weight factors of $f = 0.75, 1, \mbox{and } 1.25$. Frequency refers to the number of times $\gamma_{int}$ appears in the histogram. All simulated $\gamma_{int}$ values lie in the range predicted by Equation \ref{gammamax3}. The maximum values of $\gamma_{int}$ approaches unity, which suggests that the set of energies and transition moments that are consistent with sum rules can create an intrinsic second hyperpolarizabilities that can be close to fundamental limit ($\gamma_{int}\simeq 1$). On the other hand, the smallest $\gamma_{int}$ peaks at zero and the largest negative value approaches $-0.25$, as predicted.

In contrast to studies that determine the shape of the class of potential energy functions that optimize the intrinsic hyperpolarizabilities, which leads to $\beta_{int} < 0.708$, the Monte Carlo method yields values of $\beta_{int}$ that approach the limit.  We find that the same is true for $\gamma_{int}$.  The distribution of $\gamma_{int}$ for all three weighting factors peaks near zero. However, the distributions that are generated with larger weighting factors have larger tails near the limits, as shown in the insets of Figure \ref{fig:15}.

To better illustrate the distribution of $\gamma_{int}$ near the limits, we have plotted the histogram of $\gamma_{int}$ in the range  $[0.75, 1]$ (Figure \ref{fig:15}b) and in the range $[-0.1, -0.25]$(Figure \ref{fig:15}c) where the bin size is $0.05$.  Both of these diagrams suggest that systems whose energy difference between adjacent states gets larger for higher energy (black bars with $f=1.25$) appear more frequently near the limit.

Recall that the three-level ansatz states that when the hyperpolarizability is near the fundamental limit, the ground and two excited states dominate.  When the sum rules are applied to the three-level model of the hyperpolarizability, and the model is parameterized in terms of,
\begin{equation}\label{E}
E = \frac {e_1} {e_2} = \frac {E_{10}} {E_{20}}
\end{equation}
and
\begin{equation}\label{X}
X = \xi_{10} = \frac {x_{10}} {x_{10}^{max}},
\end{equation}
where,
\begin{equation}\label{Xn0}
x_{0n}^{max} = \sqrt{  \frac{\hbar^2 N}{2 m E_{n0}} },
\end{equation}
the hyperpolarizability is then found to be at the fundamental limit when $X=1/\sqrt[4]{3}$ and $E=0$.\cite{kuzyk00.01} The three-level ansatz and the fact that $X=1/\sqrt[4]{3}$ when the hyperpolarizability is large is found to be a universal property of all systems studied to date.\cite{kuzyk10.01}  While the fundamental limit of the off-resonant second hyperpolarizability is calculated using the three-level ansatz,\cite{kuzyk00.02} there is no extensive body of literature that supports its validity.  As we show below, the Monte Carlo simulations show that when $\gamma_{int} \rightarrow 1$, the system can be modeled with three dominant states.

\begin{figure}
\includegraphics{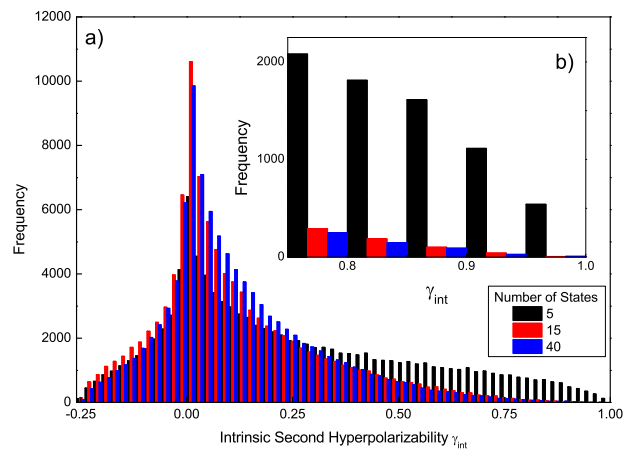}
\caption{a) Histogram of $\gamma_{int}$ for 5, 15 and 40 state models. (b) Enlarged histogram for $\gamma_{int}\in [0.75, 1]$. The bin size is $0.05$}\label{fig:states}
\end{figure}
Figure \ref{fig:states} shows the resulting distributions for Monte Carlo simulations of the intrinsic second hyperpolarizability for $5$, $10$, and $15$ states. The inset (Figure \ref{fig:states}b) focuses on the intrinsic hyperpolarizabilities in the range $[0.75, 1]$.  Clearly, when there are a fewer number of states, the frequency of $\gamma_{int}$ near the limit increases significantly: the 5-state model yields an order of magnitude higher frequency than the $15$ and $40$ state models. Also, the $15$ state systems appear with higher frequency than the $40$ state systems. This is a general trend that suggests the three level ansatz holds for the second hyperpolarizability.

\begin{figure}
\includegraphics{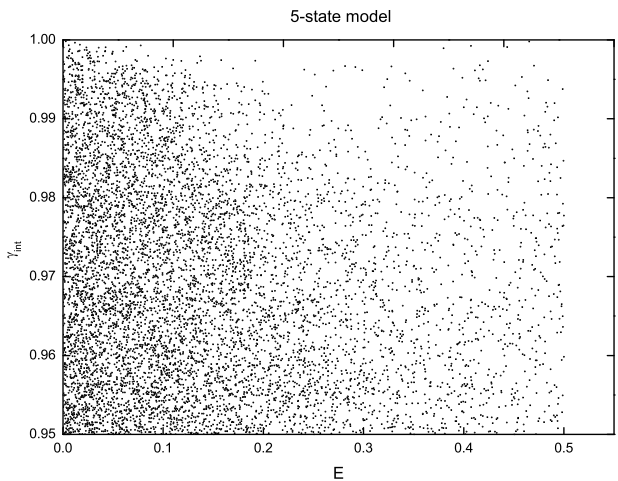}
\caption{The distribution of $\gamma_{int}$ vs. $E$. For large values of the second hyperpolarizability, the majority of $E$ values lie in the range [0, 0.1].}\label{fig:5gammaE}
\end{figure}

Figure \ref{fig:15} shows that the largest second hyperpolarizabilities are associated with greater energy spacing, suggesting that
\begin{equation}\label{limit}
\lim_{E \rightarrow 0} \gamma_{int} = 1 .
\end{equation}
Figure \ref{fig:5gammaE} plots $\gamma_{int}$ as a function of $E$ for a 5 state model and one million iterations.  When the intrinsic second hyperpolarizability approaches unity, the energy spacing most often falls in the range $E<0.1$, suggesting that $\gamma_{int}$ is the largest when $E_{20} \gg E_{10}$, the same result that is found for the first hyperpolarizability.

\begin{figure}
\includegraphics{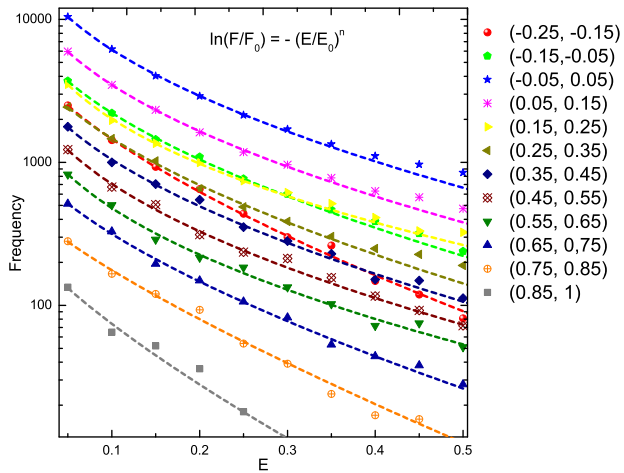}
\caption{Distribution of $E$ for different ranges of $\gamma_{int}$ for a ten-state model with 100,000 iterations. Each symbol corresponds to the specified range of $\gamma_{int}$.}\label{fig:E}
\end{figure}

To provide a more quantitative measure of the distribution of energies for various ranges of $\gamma_{int}$, we generate histograms from the data shown in Figure \ref{fig:5gammaE}.   The distribution of $E$ for a 10 state model with 100,000 iterations is shown in Figure \ref{fig:E}. Included are 12 equally spaced intervals of $\gamma_{int}$ (except for the range of $(0.85, 1)$).  The points represent the Monte Carlo data and the curves are fits to a stretched exponential model of the form,
\begin{equation}\label{fit}
\ln \left(F/F_0\right) = - \left(E/E_0\right)^n
\end{equation}
where $F_0$, $E_0$ and $n$ are fit parameters and $F$ is the frequency.  The stretched exponential was chosen as a model because it best fits the data with the fewest number of parameters.

Note that all of the curves appear to be approximately parallel to each other with the exception of the curves for $\gamma_{int}$ in the range of $(0.85,1)$, $(0.75,0.85)$, and $(-0.25,-0.15)$; which fall off more steeply as a function of $E$.  Thus, larger energy spacing is correlated with  $\gamma_{int}$ near unity or near $-0,25$, the positive and negative limits.

\begin{figure}
\includegraphics{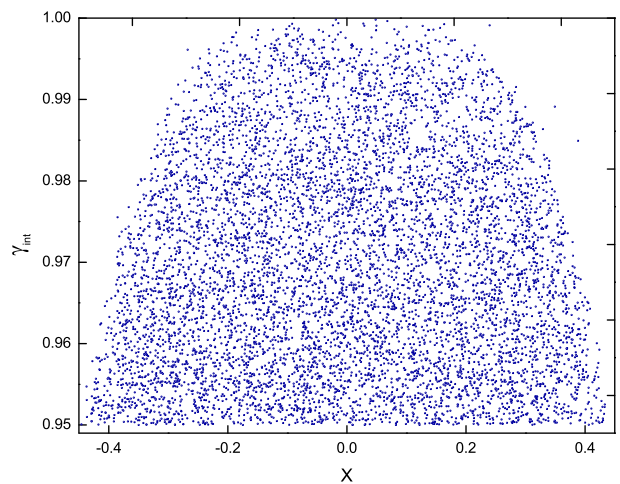}
\caption{$\gamma_{int}$ as a function of $X$ for second hyperpolarizabilities within 5\% of the limit.}\label{fig:5Xgamma}
\end{figure}

Figure \ref{fig:5Xgamma} shows $\gamma_{int}$ as a function of $X$ for a 5-state model and 100,000 runs when the second hyperpolarizabilities are larger than $0.95$.  According to Figure \ref{fig:5Xgamma}, as $\gamma_{int}$ approaches the limit, the range of $X$ becomes narrower so that for the $\gamma_{int} > 0.98$,  $X$ lies in the range  $-0.2 < X < 0.2$.

To better quantify the results, Figure \ref{fig:X,gamma} shows the distribution of $X$ for various ranges of $\gamma_{int}$ using a 10-level model.  The bin size is $0.05$ and $\gamma_{int}$ has been divided into twelve equally-spaced ranges except for $\gamma_{int} \in [0.85, 1]$. $\gamma_{int} \in [-0.25, -0.15]$ is dominated by systems with $X = \pm 1$.  For slightly smaller negative values of $\gamma_{int}$, the peak in the distribution is less than unity and the rest of the curves peak at $X=0$.
\begin{figure}
\includegraphics{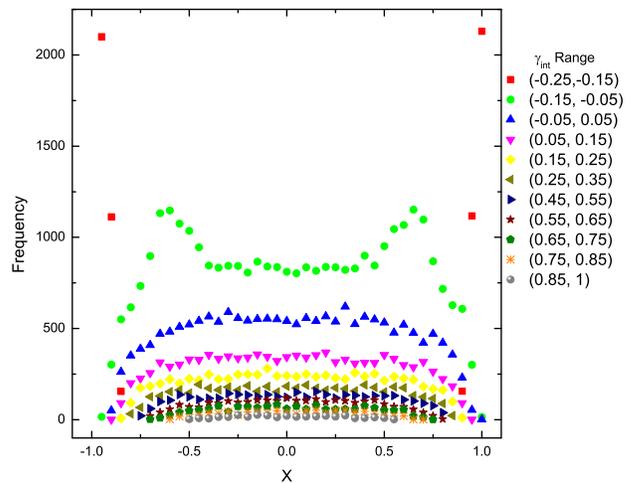}
\caption{Distribution of $X$ for different ranges of $\gamma_{int}$}\label{fig:X,gamma}
\end{figure}

While any value of $\gamma_{int}$ is attainable for any arbitrary choice of $X$, it is clear that the most likely value for the dominant transition moment is $X= \pm 1$ when $\gamma_{int} = -0.25$ and $X=0$ when $\gamma_{int} = 1$.  Recall that Figure \ref{fig:E} implies that the largest values of $\gamma_{int}$ are for $E=0$.  Interestingly, the three-level ansatz shows that when $E=0$, $X=0$ yields $\gamma_{int} = 1$ and $|X| = 1$ yields $\gamma_{int} = -0.25$.\cite{kuzyk00.02}  Thus, while the the Monte Carlo simulations include systems where many states may be contributing to the second hyperpolarizability, on average, this statistical result gives the same result as the three-level ansatz.  What remains to be investigated is the validity of the the three-level ansatz.

The three level ansatz is fundamental to the calculations of the limit of the first- and second-order nonlinear optical response.  The intrinsic hyperpolarizability $\beta_{int}$, can expressed as \cite{kuzyk08.01},
\begin{equation}\label{betaint}
\beta_{int} = {\sum _{n,m}}^\prime \beta_{int}^{n,m},
\end{equation}
where the prime indicates that the ground state is excluded in the summation. $\beta_{int}^{n,m}$ is the fractional contribution of pairs of individual states $n$ and $m$,\cite{kuzyk08.01}
\begin{equation}\label{betanm}
\beta_{int}^{n,m} = \left( \frac {3} {4} \right)^{3/4} \xi_{0n} \xi_{nm} \xi_{m0} \left(\frac {1} {e_n e_m} - \frac {2 e_m - e_n} {e_n^3} \right).
\end{equation}

The total hyperpolarizability $\beta$ can be calculated as the sum over all the fractional contributions. When the hyperpolarizability is near the fundamental limit, we find that $\beta_{int}^{n,m}$ is large for only one pair of states. This is in agreement with three level ansatz.

Expressing $\gamma_{int}$ in terms of fractional contributions of pairs of states is complicated by the fact that three excited states contribute to each term in Equation \ref{gamma}. We define $\gamma_{int}^{ijk}$ as the fractional contribution of the three states $i$, $j$ and $k$
\begin{eqnarray}\label{gammaijk}
\gamma_{int} &=& {\sum_{i}}'{\sum_{j}}'{\sum_{k}}'\gamma_{int}^{ijk} = {\sum_{i \neq j \neq k}}'\gamma_{int}^{ijk} + {\sum_{i = j \neq k}}'\gamma_{int}^{ijk} + \nonumber \\
&& {\sum_{i \neq j = k}}'\gamma_{int}^{ijk} + {\sum_{i = k \neq j}}'\gamma_{int}^{ijk} + {\sum_{i = j = k}}'\gamma_{int}^{ijk} .
\end{eqnarray}
In contrast to the hyperpolarizability, where $\beta_{int}^{nn}=0$, the second hyperpolarizability does not vanish when $i=j=k$.

The simplest approach for determining the contribution of pairs of states is the missing state analysis,\cite{Dirk89.01} where one calculates $\gamma_{int}$ in the absence of a particular set of states, $(i, j)$, denoted by $\gamma_{int}^{missing}\left(i,j\right)$.  A comparison of  $\gamma_{int}$ with $\gamma_{int}^{missing}\left(i,j\right)$ describes the joint contribution of states $i$ and $j$ to the intrinsic second hyperpolarizability. The smaller the value of $\gamma_{int}^{missing}(i,j)$, the larger the contribution of states $i$ and $j$ to $\gamma_{int}$.

\begin{figure}
\includegraphics{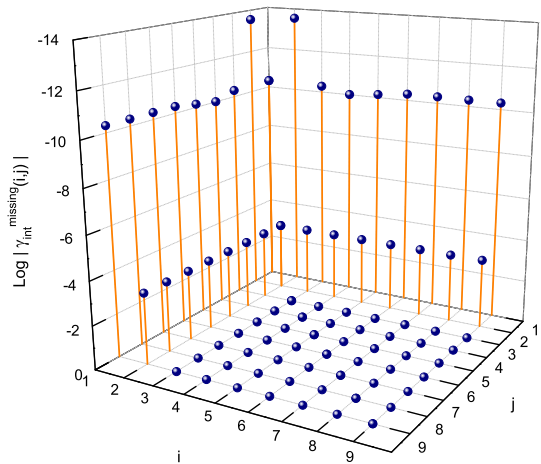}
\caption{A log plot of $|\gamma_{int}(i,j)|$ using the missing state analysis for a 10 state model and $\gamma_{int} = 0.9886$. We remark that $|\gamma_{int}(i,j)| = |\gamma_{int}(j,i)|$ so that the two smallest values in the diagram are identical and represent only one pair of states.}\label{fig:missing}
\end{figure}

Figure \ref{fig:missing} shows a logarithmic plot of the absolute value of $\gamma_{int}^{missing}(i,j)$ for a 10-state model with $\gamma_{int} = 0.9886$, the largest valued obtained in a Monte Carlo simulation.  The most important pair of states are $1$ and $2$ with $\gamma_{int}^{missing}(2,1) = 2.76444 \times 10^{-14}$. The next most important pair of states are $5$ and $1$ with $\gamma_{int}^{missing}(5,1) = 10^{-11}$. $\gamma_{int}^{missing}(8,1)$ and $\gamma_{int}^{missing}(9,1)$ are also on the order of $10^{-11}$.  Thus, the dominant two states contribute about $10^3$ times the next most important state.  The same procedure can be followed for other optimized values, and similar results are found.  It is worth mentioning that $\gamma^{missing}_{int}(i,j) \neq \gamma^{missing}_{int}(j,i)$ because the individual terms in Equation \ref{gamma} are not symmetric  when two incidences are interchanged.  Since each pair of states in Equation \ref{gamma} contribute twice, we can define
\begin{equation}\label{symmetry}
\gamma^{symmetrized}_{int}(i,j) = \gamma^{missing}_{int}(i,j) + \gamma^{missing}_{int}(j,i) ,
\end{equation}
for $i \neq j$ and restrict the sum to $i \leq j$.  We have applied such symmetrization to all values obtained with the missing state analysis.

\section{Conclusion}

Optimizing the intrinsic hyperpolarizabilities by varying the potential energy function results in a $30\%$ gap between the fundamental limit and optimized $\beta_{int}$. Experimentally measured second hyperpolarizabilities also fall far-short of the limit for a wide range of molecules, but Monte Carlo simulations show that $\beta_{int}$ can approach unity.  In the present work, we find that Monte Carlo simulations lead to values of $\gamma$ arbitrarily close to the fundamental limit, with  $-0.25 < \gamma_{int} < 1$, in agreement with analytical calculations.\cite{kuzyk00.02}  Also, the three level ansatz appears to be obeyed as shown using the missing state analysis, where the two dominant states account for over 99\% of $\gamma_{int}$.

It is important to stress that the Monte Carlo approach may lead to a broader set of transition moments and energies than are attainable with standard Hamiltonians; that is, Hamiltonians that include kinetic and potential energies of the many electrons in an atom as well as interactions with an external electromagnetic field.  As such, universal behavior that may be typical of systems described by standard Hamiltonians may not be observed in our calculations.  Never-the-less, we find that the Monte Carlo simulations, {\em on average}, are consistent with analytical results, with optimized $\gamma_{int}$ resulting when $E \approx 0$ and $|X| = 1$ for negative $\gamma$ and $X=0$ for positive $\gamma$.

This is the first study to confirm that there may be universal properties associated with the second hyperpolarizability when it is near the fundamental limit.  As such, our approach may lead to the design of better materials for third-order nonlinear-optical applications if the universal properties can be re-expressed in terms of parameters that can be varied by a synthetic chemist.

\textbf{Acknowledgements:}
 MGK thanks the National Science Foundation
(ECCS-0756936) and Wright Paterson Air Force Base for generously supporting this work.

%\bibliographystyle{\bstfile}
%\bibliography{\bibs}

\clearpage
\end{document}